\documentclass[11pt]{amsart}
\usepackage{amsmath,amssymb,amsfonts,amscd,epsfig}
\usepackage{caption}

\newcommand{\CC}{{\mathbb C}}

\newcommand{\PP}{{\mathbb P}}
\newcommand{\QQ}{{\mathbb Q}}
\newcommand{\RR}{{\mathbb R}}

\newcommand{\ii}{\mathrm{i}}

\newcommand{\dd}{\mathrm{d}}
\font \rus= wncyr10
\newcommand{\sha}{\, \hbox{\rus x} \,}
\newcommand{\zz}{\overline{z}}

\theoremstyle{plain}

\begin{document}

\title{Closed strings as single-valued open strings: \\ A genus-zero derivation}
\author{Oliver Schlotterer, Oliver Schnetz}
\address{Oliver Schlotterer\\
Max-Planck-Institut f\"ur Gravitationsphysik\\
Albert-Einstein-Institut\\
Am M\"uhlenberg 1\\
14476 Potsdam, Germany -- and -- Perimeter Institute for Theoretical Physics\\
31 Caroline Street N\\
Waterloo, ON N2L 2Y5, Canada}
\address{Oliver Schnetz\\
Department Mathematik\\
Cauerstra{\ss}e 11\\
91058 Erlangen, Germany}

\email{olivers@aei.mpg.de,
schnetz@mi.uni-erlangen.de}

\begin{abstract}
Based on general mathematical assumptions we
give an independent, elementary derivation of a theorem by Francis Brown and Cl\'ement Dupont in \cite{BD} which states that tree-level amplitudes
of closed and open strings are related through the single-valued map `sv'.
This relation can be traced back to the underlying moduli-space integrals over punctured
Riemann surfaces of genus zero. The sphere integrals $J$ in closed-string amplitudes and the disk
integrals $Z$ in open-string amplitudes are shown to obey $J = {\rm sv} \, Z$.
\end{abstract}

\maketitle


\section{Introduction}
\label{sec:1}

The study of scattering amplitudes grew into a fertile and rapidly developing research area at
the interface of particle physics, mathematics and string theory. A wealth of modern mathematical 
concepts including periods, motives and elliptic functions became a common theme in scattering
amplitudes of quantum field theory and string theory: Field-theory amplitudes encounter 
various flavors of polylogarithms via Feynman integrals, and string amplitudes are formulated in
terms of moduli-space integrals for punctured Riemann surfaces. In contrast to field theory, the
infinite number of vibration modes in string spectra introduces transcendental numbers
already into the tree level of string perturbation theory. 

More specifically, the low-energy expansions of tree-level amplitudes of both open and closed 
strings involve multiple zeta values (MZVs),
\begin{equation}
\zeta(k_1,k_2,\ldots,k_r) = \sum_{0<l_1<l_2<\ldots<l_r}^\infty l_1^{-k_1} l_2^{-k_2} \ldots l_r^{-k_r} \, , \ \ \ \ 
k_1,k_2,\ldots,k_r \in \mathbb N \, , \ \ \ \ 
k_r \geq 2 \, ,
\label{1.0}
\end{equation}
characterized by depth $r$ and weight $k_1+k_2+\ldots+k_r$. MZVs are the periods of the moduli 
space $M_{0,n}$ of $n$-punctured genus-zero surfaces \cite{Brown:2009qja}: For open strings, MZVs
arise from iterated integrals over the boundary of a disk, and closed-string tree amplitudes
are obtained from complex integration over punctures on a sphere. From the work of
Kawai, Lewellen and Tye (KLT) in 1986 \cite{Kawai:1985xq}, the sphere integrals for closed strings are known to factorize
into bilinears in disk integrals for open strings. However, the approach of KLT does not manifest
whether the `squaring procedure' for disk integrals induces any cancellations for certain classes of MZVs.
From the observations of \cite{Schlotterer:2012ny}, only the so-called single-valued subclass of MZVs
(see \cite{Bsv}) seems to persist in the final results for the sphere integrals in closed-string tree amplitudes. The 
purpose of this work is to give an elementary derivation for these conjectural selection rules. 
In fact, as will be detailed below, the closed-string amplitudes are tied to open-string 
amplitudes by the `single-valued map'.

While four-point tree-level scattering of open strings gives rise to all the Riemann zeta values
$\zeta(m) , \ 2\leq m \in \mathbb N$ in the low-energy expansion, the analogous closed-string four-point 
function only involves odd zeta values $\zeta(2k{+}1), \ k \in \mathbb N$. Here, the cancellations 
of integer powers of $\pi^{2}$ can be tracked by the closed-form representation 
of the four-point amplitudes in terms of gamma functions of the kinematic data. 
In open-string amplitudes with $n\geq 5$ external legs, in turn, the MZVs in the low-energy expansions 
include higher-depth instances and follow a more elaborate structure that can be understood in 
terms of motivic MZVs \cite{Schlotterer:2012ny} and the Drinfeld associator 
\cite{Drummond:2013vz, Broedel:2013aza}\footnote{State-of-the-art methods to compute the low-energy expansion of $n$-point disk integrals include 
matrix representations of the Drinfeld 
associator \cite{Broedel:2013aza} and recursions for off-shell versions of the disk integrals \cite{Mafra:2016mcc}
(building upon the approach via polylogarithm manipulations in \cite{Broedel:2013tta}).
For certain multiplicities $n$, explicit results are available for download via \cite{WWW},
and one can also use the connection between disk integrals and hypergeometric functions
to extract low-energy expansions, see e.g.\ \cite{Oprisa:2005wu, Stieberger:2006te, 
Boels:2013jua, Puhlfuerst:2015gta} and references therein.}.
And it took until 2012 that an all-order conjecture for the selection rules on the MZVs in closed-string 
$n$-point amplitudes could be made \cite{Schlotterer:2012ny}, based on an
experimental order-by-order inspection of the output of the KLT relations\footnote{Also see \cite{Stieberger:2009rr} for earlier work on
MZVs at weight $\leq 8$ in closed-string five- and six-point functions.
The all-order conjectures of \cite{Schlotterer:2012ny} have for instance been checked to match with the KLT relations 
up to transcendental weight 18 at five points and weight 9 at six points.}.

According to the observations in \cite{Schlotterer:2012ny}, closed-string low-energy expansions 
are conjectured to follow from the {\it single-valued} map
\cite{Bsv} of the MZVs in the disk integrals of open-string 
amplitudes \cite{Stieberger:2013wea, Stieberger:2014hba}.
A defining property of the resulting {\it single-valued MZVs}
is their descent from single-valued multiple polylogarithms at unit argument.
The procedure of F.~Brown \cite{Brown:2004} to eliminate the monodromies from harmonic polylogarithms induces
a map on MZVs which is referred to as the single-valued map sv \cite{gf, Bsv}. 
The MZVs in the low-energy expansion of $n$-point sphere integrals $J$ can be obtained from 
specific disk integrals $Z$ via the sv map:
As will be detailed below, see (\ref{3.9}), the relation conjectured by St.~Stieberger and T.~Taylor \cite{Stieberger:2014hba}
(based on results of \cite{Schlotterer:2012ny, Stieberger:2013wea})
\begin{equation}\label{JsvZ}
J={\rm sv} \, Z
\end{equation}
associates certain anti-meromorphic functions of the punctures on the sphere with cyclic orderings 
of the punctures on the disk boundary, following a Betti-deRham duality. An independent proof of
(\ref{JsvZ}) by F.~Brown and C.~Dupont was recently announced in \cite{BD}.

The proof of Brown and Dupont relies on a `motivic' version of the KLT formula (see section \ref{sec:2.4}).
This motivic KLT is proved to be closely related to the single-valued map (which is only proved to
exists in the motivic setup). Finally, the authors define `dihedral coordinates' to provide an explicit
formula which handles the poles in the Laurent expansions of string tree-level amplitudes.

Note that the proof of Brown and Dupont relies on the notion of `motivic periods'. The motivic concept
allows one to lift integrals from pure numbers (or functions) to objects in algebraic geometry. These
objects contain the initial data of the integral (the form and the cycle) and provide a restricted set
of transformations in algebraic cohomology. This bypasses notoriously difficult issues with transcendentality:
While in many cases it is easy to see that certain numbers (such as MZVs) are related by equations, it is much harder to prove
that a pair of numbers can never be related by a class of operations.

The notion of `motivic periods' is a mathematically beautiful and deep construction which
may not be readily accessible to physicists. The main claims relating motivic periods with 
pure numbers are:
\begin{itemize}
\item Many properties of pure numbers can only be proved in the motivic setup (like e.g.\ the existence of a weight grading of MZVs).
\item The motivic setup is conjectured to be fully equivalent (isomorphic) to the pure number setup.
\item Any explicit relation which is derived within the motivic setup is also (proved to be) true in the pure number context.
\end{itemize}

Because in (\ref{JsvZ}) the objects $J$ and $Z$ are related by the single-valued map, the result 
can only be proved in the motivic context. Therefore, the mathematically beautiful proof of Brown 
and Dupont inevitably uses more advanced mathematics which may be somewhat less accessible to physicists.

In this work, we will deliver an elementary inductive derivation (a proof under general mathematical assumptions) 
that sphere integrals are single-valued versions of disk integrals and, equivalently, that closed-string tree-level 
amplitudes are single-valued open-string amplitudes. The driving force for the derivation is the notion of single-valued integration
\cite{gf} along with its properties that originate from motivic algebraic
geometry \cite{Bsv,BrownMotPer}. The Betti-deRham duality between
the anti-meromorphic factors in the integrands on the sphere and integration cycles on the disk boundary will arise naturally from the Stokes theorem.

The sv relations between disk and sphere integrals can be applied to closed strings in the supersymmetric, 
heterotic and bosonic theories \cite{Stieberger:2014hba} and triggered several directions of follow-up research. 
For instance, single-valued open-string amplitudes govern amplitude relations mixing gauge and gravitational states 
of the heterotic string \cite{Schlotterer:2016cxa} as well as the recent double-copy description of bosonic 
and heterotic strings \cite{Azevedo:2018dgo}. 
Moreover, the appearance of single-valued MZVs in the sigma-model approach to effective gauge interactions of type-I 
and heterotic strings has been studied in \cite{Fan:2017uqy}. The derivation in this work and the proof of Brown and Dupont will place these results on firm grounds without
the need to rely on a conjectural status for the key relations (\ref{JsvZ}) between sphere integrals and single-valued disk integrals.

The derivation in this article is not a proof in a full mathematical sense for the following three reasons:

Mostly, one has to keep in mind that the single-valued map is only defined to exist in a `motivic' framework.
Because the singular divisors in the disk and sphere integrals are not normal crossing, it is a non-trivial step to
set up a motivic theory for these objects. Alternatively, one can assume standard transcendentality conjectures for the related integrals.

Moreover, we use three natural properties of the single-valued map in section \ref{sec:3.1}. These properties are thoroughly tested.
Properties (i) and (ii) are proved or mostly proved in the stated literature. Property (iii) is proved in the text using a standard
property of the `$f$-alphabet' which may not be fully proved in the mathematical literature (the $f$-alphabet has not yet drawn much
attention in mathematics).

Finally, we use the existence of a subtraction scheme whose existence we do not prove here.
In tree-level string theory, the purpose of subtraction schemes is to capture the kinematic poles in disk and sphere integrals.
These poles have already been investigated from various different perspectives 
\cite{Mafra:2011nw, Broedel:2013tta, Mafra:2016mcc}. Moreover, subtractions schemes are extensively studied 
in the much more complicated case of quantum field theory (see e.g.\ \cite{BKangles,numfunct}).
Further, note that in \cite{BD} Brown and Dupont prove the existence of this subtraction scheme in full mathematical rigor.
So, we considered it more beneficial to provide the reader with explicit examples in appendix A and 
refer to \cite{BD} for the full proof.

In spite of these restrictions we informally use the word `proof' in this article.


\section{Reviewing the bases and relations of disk and sphere integrals}
\label{sec:2}

In this section, we review the classes of disk and sphere
integrals that are related through the sv map. These moduli-space integrals encode
the low-energy regime of string tree-level amplitudes through 
their series expansion in the dimensionless Mandelstam invariants
\begin{equation}
s_{ij} := 2\alpha' k_i\cdot k_j = s_{ji} \ , \ \ \ \ \ \ s_{ij} \in \mathbb R \, ,
\label{2.0}
\end{equation}
where $\alpha'$ denotes the inverse string tension. The external
momenta $k_{i}$ are Lorentz vectors referring to massless external 
states $i=1,2,\ldots,n$ of an $n$-point amplitude subject to $k_i^2=0$ 
and momentum conservation $\sum_{i=1}^n k_i=0$. These kinematic
constraints imply
\begin{equation}
 s_{i,i}=0,\quad \quad \sum_{i=1}^n s_{ij} = 0 \ \forall\ j=1,2,\ldots,n\, ,
\label{2.0A}
\end{equation}
so that only $\frac{n}{2}(n{-}3)$ Mandelstam invariants are independent.


\subsection{Four-point integrals: an inviting example}
\label{sec:2.1}

The simplest appearance of MZVs in string perturbation theory
occurs in the four-point tree amplitude of open strings. After peeling
off suitable kinematic factors, the amplitude boils down to the disk integral
\begin{equation}
Z_{\rm 4pt}  := \int^1_0 \frac{ {\rm d} z }{z} z^{s_{12}} (1-z)^{s_{23}} = \frac{ \Gamma(s_{12}) \Gamma(1+s_{23}) }{\Gamma(1+s_{12}+s_{23})} 
\label{2.1}
\end{equation}
and its permutations w.r.t.\ the external momenta. The $\alpha'$-expansion of the integral $Z_{\rm 4pt}$ 
-- i.e.\ the simultaneous series expansion in the dimensionless $s_{ij}$ variables (\ref{2.0}) -- follows 
from the $\Gamma$-function 
identity $\log \Gamma(1+x) = -\gamma x + \sum_{k=2}^{\infty} \frac{ \zeta(k) }{k}(-x)^k$,
\begin{align}
Z_{\rm 4pt} &= \frac{1}{s_{12}} \exp\Big( \sum_{k=2}^\infty \frac{ \zeta(k) }{k} (-1)^k \big[ s_{12}^{k}+s_{23}^k - (s_{12}+s_{23})^k \big] \Big) \notag \\
&= \frac{1}{s_{12}} - \zeta(2) s_{23} + \zeta(3) s_{23}(s_{12}+s_{23}) +O(\alpha'^3) \, ,
\label{2.2}
\end{align}
and involves all Riemann zeta values (while the Euler Mascheroni constant $\gamma$ cancels).

The simplest appearance of MZVs in a closed-string setup is the following complex integral in the four-point tree amplitude
\begin{equation}
J_{\rm 4pt} := \frac{1}{\pi} \int \limits_{\mathbb C}
\frac{ {\rm d}^2 z }{z \bar z(1-\bar z)} |z|^{2s_{12}} |1-z|^{2s_{23}} = \frac{ \Gamma(s_{12}) \Gamma(1+s_{23}) \Gamma(1+s_{13}) }{\Gamma(1-s_{12})\Gamma(1-s_{23})\Gamma(1-s_{13})}  \, ,
\label{2.3}
\end{equation}
where $\zz$ is the complex conjugate of $z=x+\ii y$ and $\dd^2z :=\dd x \, \dd y$. The $\alpha'$-expansion takes a particularly
symmetric form in terms of $s_{13}=-s_{12}-s_{23}$,
see (\ref{2.0}),
\begin{align}
J_{\rm 4pt} &= \frac{1}{s_{12}}\exp\Big( {-}2 \sum_{k=1}^{\infty} \frac{ \zeta(2k{+}1) }{2k{+}1}  \big[ s_{12}^{2k+1}+s_{23}^{2k+1} + s_{13}^{2k+1}  \big] \Big) \notag \\
&= \frac{1}{s_{12}} + 2 \zeta(3) s_{23} (s_{12}+s_{23}) +O(\alpha'^4) \, ,
\label{2.4}
\end{align}
and the first line of (\ref{2.4}) manifests the cancellation of even Riemann-zeta values.


\subsection{The integrals for $n$ points}
\label{sec:2.2}

The above four-point integrals fall into the following general classes of $n$-point disk integrals $Z(\tau |\rho)$
and sphere integrals $J(\tau |\rho)$,
\begin{align}
Z(\tau| \rho) &:= \! \! \! \! \!  \! \! \! \! \! \int\limits_{-\infty \leq z_{\tau(1)} \leq z_{\tau(2)}\leq\ldots\leq z_{\tau(n)}\leq\infty}  \! \! \! \! \!  \! \! \! \! \!   \frac{ \dd z_1 \, \dd z_2 \, \ldots \, \dd z_n }{{\rm vol} \, {\rm SL}_2(\mathbb R)} \, \frac{(-1)^{n-3}\, \prod_{1\leq i<j\leq n} |z_{i,j}|^{s_{ij} } }{  z_{\rho(1),\rho(2)} z_{\rho(2),\rho(3)}
\ldots z_{\rho(n-1),\rho(n)}z_{\rho(n),\rho(1)} },
\label{poles01} \\
J(\tau| \rho) &:= \int\limits_{\mathbb C^n}    \frac{ \dd^2 z_1 \, \dd^2 z_2 \, \ldots \, \dd^2 z_n }{\pi^{n-3} \, {\rm vol} \, {\rm SL}_2(\mathbb C)} \, \frac{ \prod_{1\leq i<j\leq n} |z_{i,j}|^{2s_{ij} } }{  (z_{\rho(1),\rho(2)}z_{\rho(2),\rho(3)}
\ldots z_{\rho(n),\rho(1)})  \, (\bar z_{\tau(1),\tau(2)} \bar z_{\tau(2),\tau(3)} \ldots \bar z_{\tau(n),\tau(1)})}  \, ,
\label{poles02}
\end{align}
where $z_{i,j}:=z_i-z_j$. Both types of integrals 
are indexed by two permutations $\rho ,\tau \in S_n$ of
the legs $\{1,2,\ldots,n\}$. The absolute value in the integrand 
$\prod_{1\leq i<j\leq n} |z_{i,j}|^{s_{ij}}$
of (\ref{poles01}) ensures that only positive numbers are raised to the power of $s_{ij}$, regardless of the 
integration domain characterized by $z_{\tau(i)} < z_{\tau(i+1)}$.

The inverse factor of ${\rm vol} \, {\rm SL}_2(\mathbb R)$ in the disk integrals (\ref{poles01}) is implemented by 
dropping three integrations over any $z_i,z_j,z_k$ (with $i,j,k \in \{1,2,\ldots,n\}$), inserting $|z_{i,j} z_{i,k}z_{j,k}|$ and fixing 
$(z_i,z_j,z_k)\rightarrow (0,1,\infty)$. Its analogue $({\rm vol} \, {\rm SL}_2(\mathbb C))^{-1}$
in the sphere integral (\ref{poles02}) instructs to insert $|z_{i,j} z_{i,k}z_{j,k}|^2$.
The limit $z_k\rightarrow \infty$ is non-singular by the Mandelstam identity (\ref{2.0A}) and the choice of cyclic `Parke--Taylor' 
denominators in (\ref{poles01}) and (\ref{poles02}). 

Note that the four-point integrals (\ref{2.1}) and (\ref{2.3}) can be recovered from the general definition via
\begin{equation}
Z_{\rm 4pt} =  - Z(1,2,3,4|1,2,4,3) \ , \ \ \ \ \ \ J_{\rm 4pt} =  - J(1,2,3,4|1,2,4,3)
\label{2.5}
\end{equation}
after fixing $(z_1,z_3,z_4) \rightarrow (0,1,\infty)$ and identifying $z_2\rightarrow z$.
The low-energy expansions (\ref{2.2}) and (\ref{2.4}) of the four-point integrals generalize as follows
to higher multiplicity: The $n$-point integrals (\ref{poles01}) and (\ref{poles02}) admit a Laurent expansion
in the dimensionless Mandelstam invariants (\ref{2.0}) of the form \cite{Mafra:2011nw, Schlotterer:2012ny, Broedel:2013aza} 
\begin{align}
Z(\tau| \rho) &=  p_{3-n}(\tau| \rho) + \zeta(2) p_{5-n}(\tau| \rho) + \zeta(3) p_{6-n}(\tau| \rho) 
+ \zeta(4) p_{7-n}(\tau| \rho) + O(\alpha'^{8-n}) \, ,
\label{Lau01}
\end{align}
where $p_{k}(\tau|\rho)$ are Laurent polynomials in $s_{i\ldots j}= \alpha'(k_i+\ldots+k_j)^2$ of homogeneity 
degree $k$ with rational coefficients. The $\alpha'$-expansion of $J(\tau| \rho) $ follows the same structure:
Equation (\ref{JsvZ}) translates the leading low-energy orders (\ref{Lau01}) of the disk integrals into
\begin{align}
J(\tau| \rho) &=  p_{3-n}(\tau| \rho) +2\zeta(3) p_{6-n}(\tau| \rho) + O(\alpha'^{8-n}) \, ,
\label{Lau02}
\end{align}
cf.\ (\ref{3.6}), with the same degree-$k$ Laurent polynomials $p_{k}(\tau| \rho)$ in $s_{ij}$ 
as seen in the $\alpha'$-expansion (\ref{Lau01}) of the disk integrals.

By the results of refs.\ \cite{Mafra:2011nv, Broedel:2013tta}, the $n$-point tree-level amplitudes of open and 
closed superstrings are expressible in terms of the integrals (\ref{poles01}) and (\ref{poles02}), also see 
\cite{Huang:2016tag, Azevedo:2018dgo} for analogous statements on bosonic and heterotic strings.


\subsection{Relations of disk and sphere integrals}
\label{sec:2.3}

One can infer from the right-hand sides of (\ref{poles01})  and (\ref{poles02}) that the disk and sphere
integrals $Z(\tau |\rho)$ and $J(\tau |\rho)$ only depend on the cyclic equivalence class of the 
permutations $\tau,\rho$. The cyclic denominators manifest that
\begin{equation}
Z(\tau|1,2,3,\ldots,n) = Z(\tau|2,3,\ldots,n,1) \, , \ \ 
J(\tau|1,2,3,\ldots,n) = J(\tau|2,3,\ldots,n,1)  \ \forall \ \tau \in S_n\, ,
\label{2.11}
\end{equation}
and the same is true for the first entry of the sphere integrals [by reality $J(\tau |\rho)=J(\rho|\tau)$].
Also, the integration domain of the disk integrals (\ref{poles01}) is cyclically invariant
\begin{equation}
Z(1,2,3,\ldots,n|\rho) = Z(2,3,\ldots,n,1|\rho)  \ \forall \ \rho \in S_n \, .
\label{2.12}
\end{equation}
Still, the number $(n{-}1)!$ of cyclically inequivalent
permutations in $S_n$ overcounts the number of
inequivalent disk and sphere integrals: Different choices of the cyclic denominators are related via integration-by-parts 
relations which lead to a basis of $(n{-}3)!$ inequivalent permutations of $(z_{1,2} z_{2,3}\ldots z_{n,1})^{-1}$. 
For disk integrals, dropping total derivatives w.r.t.\ the punctures yields \cite{Broedel:2013tta}
\begin{equation}
\sum_{j=2}^{n-1} k_1 {\cdot}(k_2{+}k_3{+}\ldots{+}k_j) Z(\tau|2,3,\ldots,j,1,j{+}1,\ldots,n{-}1,n) =0 \ \forall \ \tau \in S_n \, ,
\label{2.13}
\end{equation}
and the same relations hold for both entries of the sphere integrals.
Since the first entry of the disk integrals (\ref{poles01}) refers to an integration cycle 
$-\infty \leq z_{\tau(1)} \leq z_{\tau(2)}\leq\ldots\leq z_{\tau(n)}\leq\infty$ rather than a choice
of integrand, i.e.\ $Z(\tau |\rho)\neq Z(\rho |\tau)$, monodromy properties of the Koba--Nielsen factor 
$\prod_{1\leq i<j\leq n} |z_{i,j}|^{s_{ij} }$ yield \cite{BjerrumBohr:2009rd, Stieberger:2009hq}
\begin{equation}
\sum_{j=2}^{n-1} \sin\big[2\pi \alpha' k_1 {\cdot}(k_2{+}k_3{+}\ldots{+}k_j) \big] Z(2,3,\ldots,j,1,j{+}1,\ldots,n{-}1,n|\rho) =0
 \ \forall \ \rho \in S_n \, .
\label{2.14}
\end{equation}
The combinatorics of these monodromy relations follow the structure of (\ref{2.13}) except for the 
promotion of the coefficients $k_1 {\cdot}(k_2{+}k_3{+}\ldots{+}k_j)$ to a trigonometric function. Hence,
permutations of (\ref{2.14}) leave $(n{-}3)!$ independent integration cycles \cite{BjerrumBohr:2009rd, Stieberger:2009hq}. 

By combining permutations of (\ref{2.13}) and (\ref{2.14}), 
the moduli-space integrals $Z(\tau |\rho)$ and $ J(\tau |\rho)$ can be expressed in a basis of $(n{-}3)! \times (n{-}3)!$
elements. For both entries, one can fix legs $n{-}1,n,1$ in adjacent positions and take $\rho=1,\beta,n{-}1,n$ 
with permutations $\beta\in S_{n-3}$ of $\{2,3,\ldots,n{-}2\}$ as a convenient basis choice. These relations can 
be understood in the framework of intersection theory, where $(n{-}3)!$ arises as the dimension of twisted 
homologies and cohomologies \cite{Mizera:2017cqs, Mizera:2017rqa}. 


\subsection{Kawai--Lewellen--Tye relations}
\label{sec:2.4}

Using the representations of the four-point integrals (\ref{2.1}) and (\ref{2.3}) in terms of $\Gamma$ functions,
one can observe via $\sin(\pi x) = \frac{ \pi }{ \Gamma(1-x) \Gamma(x) }$ that
\begin{equation}
J(1,2,3,4|1,2,4,3) = - \frac{ 1}{\pi}  Z(1,2,3,4|1,2,4,3) \sin(\pi s_{12}) Z(1,2,4,3|1,2,3,4)\, .
\label{2.15}
\end{equation}
This is the simplest instance of the KLT relations \cite{Kawai:1985xq} between sphere integrals and bilinears in
disk integrals which can be derived by suitable deformations of the complex integration contours.
Their generalizations to $n$-points do not depend on the Parke-Taylor denominators
in the integrand of $J(\tau|\rho)$ and may be described in terms of a $(n{-}3)! \times (n{-}3)!$ KLT matrix
$S_{\alpha'}(\sigma | \beta)_1$ \cite{Kawai:1985xq, Bern:1998sv, BjerrumBohr:2010hn}
\begin{equation}
J(\tau|\rho) = \sum_{\sigma,\beta \in S_{n-3}}  Z(1,\sigma,n,n{-}1|\tau) S_{\alpha'}(\sigma | \beta)_1  Z(1,\beta,n{-}1,n|\rho)\, .
\label{2.16}
\end{equation}
The KLT matrix $S_{\alpha'}(\sigma | \beta)_1$ is indexed by permutations
$\sigma,\beta \in S_{n-3}$ of $\{2,3,\ldots,n{-}2\}$ and admits a recursive definition \cite{BjerrumBohr:2010hn,Carrasco:2016ldy}
\begin{align}
S_{\alpha'}(2|2)_1 &= - \frac{ 1}{\pi} \sin(\pi s_{12})  = - \frac{ 1}{\pi} \sin(2\pi \alpha' k_{1}\cdot k_{2})  \label{2.17}\\
S_{\alpha'}(A,j|B,j,C)_1 &= - \frac{ 1}{\pi} \sin\big(2\pi \alpha' k_j \cdot (k_1+k_B) \big)  S_{\alpha'}(A|B,C)_1\, .  \notag
\end{align}
Here, we are employing the notation $A=a_1a_2\ldots a_p$ and $B=b_1 b_2\ldots b_q$ for words of 
length $p,q\geq 0$ composed of external-state labels $a_{i}$ and $b_j$ as their letters.
We also use $k_B=\sum_{j=1}^q k_{b_j}$ for the overall momentum associated with the word $B=b_1 b_2\ldots b_q$.
The recursive step in (\ref{2.17}) removes the last leg $j$ in the first entry of $S_{\alpha'}(\cdot | \cdot)_1 $ which
is not necessarily in the last position in the second entry. The subscript of $S_{\alpha'}(\sigma | \beta)_1$ indicates
that the entries in (\ref{2.17}) depend on both $k_1$ and the momenta 
$k_2,k_3\ldots,k_{n-2}$ associated with the permutations $\sigma, \beta$.

Similar to the integration-by-parts and monodromy relations (\ref{2.13}) and (\ref{2.14}), the KLT 
relations (\ref{2.16}) can be elegantly understood in terms of intersection theory \cite{Mizera:2017cqs} where they follow from
the twisted period relations \cite{ChoMatsumoto}.

The permutations $1,\sigma,n,n{-}1$ and $1,\beta,n{-}1,n$ in (\ref{2.16}) reflect a particular
basis choice of twisted homologies that is tailored to simplify the KLT matrix (\ref{2.17}):
The three legs $1,n{-}1,n$ are kept in adjacent positions, and the sets of integration cycles
for $Z(1,\sigma,n,n{-}1|\tau)$ and $Z(1,\beta,n{-}1,n|\rho)$ in (\ref{2.16}) are related through the
transposition $n{-}1 \leftrightarrow n$. With this choice of bases, the entries of $S_{\alpha'}(\sigma | \beta)_1$
do not depend on $k_{n-1}$ or $k_n$.

Given the $\alpha'$-expansion of the disk integrals $Z(\tau|\rho) $, the KLT relations (\ref{2.16}) in principle
determine the analogous expansion of $J(\tau|\rho)$. However, already the four-point example (\ref{2.15})
reveals the shortcoming of the KLT relations that both of its ingredients $Z(\tau|\rho) $ and $S_{\alpha'}(\sigma | \beta)_1$
carry spurious contributions of $\zeta(2k) , \ k \in \mathbb N,$ which are absent in the final result (\ref{2.4}).

At $n\geq 5$ points, similar cancellations have been observed \cite{Schlotterer:2012ny} by inserting explicit $\alpha'$-expansions
of disk integrals into KLT formulae equivalent to (\ref{2.16}).
In the following we will not use the KLT relations. We rather give a general proof that the
observed patterns of MZVs in sphere integrals are governed by the single-valued map.

Note that, in contrast to our approach, the proof of (\ref{JsvZ}) in \cite{BD} uses a motivic version of the KLT relations.


\section{The main result}
\label{sec:3}


\subsection{Single-valued iterated integrals and single-valued MZVs}
\label{sec:3.1}

The notion of single-valued (motivic\footnote{The single-valued map is only proved to exist in the motivic context \cite{Bsv}.}) MZVs is based on the representation of generic MZVs (\ref{1.0}) 
in terms of multiple (harmonic) polylogarithms at unit argument (see \cite{Chen} for the general definition of iterated integrals $I$)
\begin{align}
I(0,a_1a_2\ldots a_w,z) &= \int^z_0 \frac{ {\rm d} t}{t-a_w} I(0,a_1a_2\ldots a_{w-1},t) \, , \ \ \ \ I(0,z) = 1\, ,
\label{3.1} \\
\zeta(n_1,n_2,\ldots,n_r) &= (-1)^r I(0,\underbrace{100\ldots0}_{n_1}\underbrace{100\ldots0}_{n_2}\ldots
\underbrace{100\ldots0}_{n_r},1)\, ,
\label{3.2}
\end{align}
where $z \in \mathbb C$. For each choice of $a_1,a_2,\ldots,a_w \in \{0,1\}$, a construction by F.~Brown 
\cite{Brown:2004} provides a unique single-valued iterated integral ${\mathcal I} (0,a_1a_2\ldots a_w,z)$.
The latter can be considered as iteratively performing `single-valued integrations' from the base point 0
to $z$ in complete analogy to the analytic integration in (\ref{3.1}).

In such single-valued multiple polylogarithms the monodromies of (\ref{3.1}) around $t=0,1,\infty$ are annihilated 
by anti-holomorphic admixtures, e.g.
\begin{align}
{\mathcal I}(0,1,z) &=I(0,1,z){+}I(0,1,\bar z)  \, , \ \ \ \ {\mathcal I}(0,10,z) = I(0,10,z) {+} I(0,0,z)I(0,1,\bar z) {+} I(0,01,\bar z)\, ,
\label{3.3}
\\
{\mathcal I}(0,100,z) &= I(0,100,z) + I(0,00,z) I(0,1,\bar z)+ I(0,0,z) I(0,01,\bar z) + I(0,001,\bar z)\, .
\notag
\end{align}
While the holomorphic differentials $\frac{\partial}{\partial z}$ of $I(0,\ldots,z)$ are preserved by the ${\mathcal I}(0,\ldots,z)$,
the general connection between $\mathcal I$ and $I$ is more complicated than suggested in the above examples (a Maple implementation is \cite{Hyperlogproc}).
By analogy with (\ref{3.2}), single-valued MZVs (and the corresponding single-valued map sv) are defined as 
single-valued multiple polylogarithms at unit argument \cite{gf, Bsv},
\begin{align}\label{zetasv}
\zeta_{\rm sv}(n_1,n_2,\ldots,n_r) &= (-1)^r {\mathcal I}(0,\underbrace{100\ldots0}_{n_1}\underbrace{100\ldots0}_{n_2}\ldots
\underbrace{100\ldots0}_{n_r},1)\\
{\rm sv}: \ \zeta(n_1,n_2,\ldots,n_r) &\rightarrow
\zeta_{\rm sv}(n_1,n_2,\ldots,n_r) \, .
\notag
\end{align}
At the level of Riemann zeta values, single-valued MZVs (\ref{zetasv}) take the simple form
\begin{equation}
\zeta_{\rm sv}(2k) = 0 \, , \ \ \ \ \zeta_{\rm sv}(2k{+}1) = 2 \zeta(2k{+}1) \, ,
\label{3.6}
\end{equation}
while higher-depth instances such as
\begin{equation}
\zeta_{\rm sv}(3,5) = -10\zeta(3)\zeta(5) \, , \ \ \ \ \zeta_{\rm sv}(3,5,3) = 2\zeta(3,5,3) - 2 \zeta(3)\zeta(3,5) -10\zeta(3)^2\zeta(5)  \, ,
\label{3.7}
\end{equation}
are most conveniently understood in terms of the $f$-alphabet for MZVs \cite{Bsv,Brown:falphabet}.

In the $f$-alphabet (motivic) iterated integrals become words in some alphabet which reflects the number-theoretical contents 
of the iterated integral. The $f$-alphabet exists for arbitrary $a_1,\ldots,a_w\in \CC $, in which case the iterated integrals (\ref{3.1}) are hyperlogarithms.
(Single) logarithms are primitive, i.e.\ they are represented by a single letter (of weight one). The product becomes shuffle $\sha$,
and there is some admixture of polynomial
type from pure periods (integrals without boundary which in the case of hyperlogarithms are polynomials in $2\pi\ii$).

Iterated integrals in several analytic variables are represented by words with purely analytic 
letters\footnote{In the context of quantum field theory, iterated integrals
with non-analytic letters also play a prominent role \cite{numfunct,GSVH}. Handling these objects is more complicated. Here, we only need the straightforward analytic case.}.
In an $f$-alphabet with purely analytic letters the sv map on a word $w$ is given by
\begin{equation}\label{svf}
\mathrm{sv}\,w=\sum_{w=uv}\overline{\tilde u}\sha v\, ,
\end{equation}
where $\tilde u$ is $u$ in reversed order (and $\overline{\bullet}$ is complex conjugation).
Moreover, sv $2\pi\ii=0$. In physical terminology the $f$-alphabet can be considered as a complete symbol \cite{SchnetzAlgIntTalks}. In particular, the conversion into the
$f$-alphabet has trivial kernel, so that no information is lost when one uses the $f$-alphabet.

In pure mathematics the sv map exists as evaluation of `deRham' periods in a very general motivic context. Here, we only use sv as the map
\begin{equation}
\text{sv:} \ I(0,a_1a_2\ldots a_w,z)\mapsto{\mathcal I}(0,a_1a_2\ldots a_w,z)
\label{3.13}
\end{equation}
(which is consistent with (\ref{zetasv})). In general, there exist relations between iterated integrals (e.g.\ for MZVs).
A priori it is unclear (surprising even) that the map sv is well-defined (i.e.\ it is consistent with all relations).
However, the sv-map on $I(0,a_1a_2\ldots a_w,z)$ can be proved to have the following three natural properties:
\begin{enumerate}
\item[(i)] The sv-map is well-defined.
\item[(ii)] The sv-map commutes with evaluation.
\item[(iii)] The sv-map extends to several (analytic) variables. 
I.e.\ $\mathcal I(0,a_1a_2\ldots a_w,z)$ is single-valued in all variables $a_1,a_2,\ldots,a_w,z$ of its letters.
\end{enumerate}
These results have a deep origin in motivic algebraic geometry.
The Ihara action \cite{Ihara} plays a major role in the proof of property (i) for iterated integrals.
Property (i) is theorem 1.1 in \cite{Bsv} and property (ii) in the context 
of multiple polylogarithms is corollary 5.4 in \cite{Bsv}.
More on the evaluation of hyperlogarithms at special values of the arguments can be found in \cite{BrownMotPer}.

Property (iii) can be proved in the $f$-alphabet \cite{Brown:falphabet}:
\begin{proof}[Proof of (iii)]
In the general hyperlogarithmic context, monodromies can be expressed in terms of an `infinitesimal' object, $\mathcal M=\exp(m)$,
Here, $m$ can be considered as picking the part of the monodromy which is proportional to $2\pi\ii$. Note that $m$ is a derivative (i.e.\ it obeys the Leibniz rule).
In the $f$-alphabet, $m$ is obtained from the first letter on the Betti side (here, the left-hand side) \cite{SchnetzAlgIntTalks},
\begin{equation}
m(aw)=m(a)w\, ,
\end{equation}
where $a$ is a letter and $w$ is a word.

Expressions with trivial monodromy lie in the kernel of $m$. 

For hyperlogarithms, the only functions represented by single letters in the $f$-alphabet are logarithms (all logarithms are `primitives' of weight one).
Hence, only words with logarithms (like $I(0,a_1,z)=\log(1-z/a_1)$) as first letters contribute to the differential monodromy $m$.
For such a logarithm the differential monodromy around $z=a_1$ is $2\pi\ii$. The complex conjugate letter $\log(1-\zz/\overline{a}_1)$ has
differential monodromy $-2\pi\ii$ around $z=a_1$ (this also remains true if one considers the monodromy of the variable $a_1$ around a fixed value of $z$).
From this we conclude that in the $f$-alphabet for hyperlogarithms single-valuedness means that all words not beginning in constants
come in pairs with complex conjugate first letters.

For a letter $a$ we define $\partial_a aw=w$ (clipping off the first Betti letter)
and $\partial_a bw=0$ if $b\neq a$. Note that $\partial_a$ is a differential with respect to the shuffle product.
Because of the monodromy property of the $f$-alphabet, the proof of property (iii) reduces to showing that
\begin{equation}
\partial_a \mathrm{sv}\,w=\partial_{\overline{a}} \mathrm{sv}\,w
\end{equation}
for all words $w$ and all letters $a$ (with complex conjugate $\overline{a}$).
From (\ref{svf}) we have
\begin{equation}
\partial_a \mathrm{sv}\,w
=\sum_{w=uv} \big[ (\partial_a\overline{\tilde u})\sha v+\overline{\tilde u}\sha \partial_a v \big]
=\sum_{w=uv}\overline{\tilde u}\sha \partial_a v\, .
\end{equation}
Likewise,
\begin{equation}
\partial_{\overline{a}}\mathrm{sv}\,w=\sum_{w=uv}(\overline{\partial_a\tilde u})\sha v\, .
\end{equation}
Both expressions on the right-hand sides are equivalent to
\begin{equation}
\sum_{w=uav}\overline{\tilde u}\sha v
\end{equation}
which completes the proof. \end{proof}

Note that property (iii) means that single-valued integration with respect to any variable of a single-valued 
iterated integral is single-valued in all variables. A priori, this property of `single-valued integration' is as mysterious as properties (i) and (ii).
Single-valued integration was originally introduced by F.~Brown using generating functions \cite{Bsv}. 
In practice, it is more convenient to use a bootstrap algorithm first defined in \cite{gf}. A practical and fully 
general approach uses a commutative hexagon \cite{numfunct,GSVH}.

Also note that property (iii) relates $\mathcal I(0,a_1a_2\ldots a_w,z)$ to the single-valued multiple polylogarithms
in more than one variable constructed in \cite{Broedel:2016kls, DelDuca:2016lad}. 


\subsection{The claim}
\label{sec:3.2}

The single-valued map (\ref{3.6}) of Riemann zeta values relates the
four-point integrals of section \ref{sec:2.1} at the level of their
$\alpha'$-expansions in (\ref{2.2}) and (\ref{2.4}),
\begin{equation}
 J(1,2,3,4|1,2,4,3)= {\rm sv} \, Z(1,2,3,4|1,2,4,3)  \, ,
\label{3.8}
\end{equation}
where ${\rm sv}$ is understood
on the expansion in the parameters $s_{ij}$. By $\zeta_{\rm sv}(2k) = 0$, the sv map rationalizes the trigonometric functions
$\sin(\pi s_{ij}) = \pi s_{ij} \exp\big( {-}2 \sum_{k=1}^{\infty} \frac{ \zeta(2k) }{2k} s_{ij}^{2k} \big)$ 
in the monodromy relations (\ref{2.14}), ${\rm sv} \,  \sin(\pi s_{ij}) /\pi  = s_{ij} $. Hence, the observation 
(\ref{3.8}) extends to all four-point disk and sphere integrals of the general 
form (\ref{poles01}) and (\ref{poles02}).

The general conjecture of St.~Stieberger and T.~Taylor we want to prove in this work concerns the striking
connection between $n$-point disk and sphere integrals in (\ref{poles01}) and (\ref{poles02}) via \cite{Stieberger:2014hba}
\begin{equation}
J(\tau | \rho) = {\rm sv} \, Z(\tau | \rho) \ \forall \ \tau,\rho \in S_{n} \, .
\label{3.9}
\end{equation}
This relation identifies sphere integrals $J$ as single-valued disk integrals ${\rm sv}\, Z$, where the anti-meromorphic part
$(\zz_{\tau(1),\tau(2)} \ldots \zz_{\tau(n),\tau(1)})^{-1}$ of the sphere integrand reflects the ordering of the integration cycle
$-\infty \leq z_{\tau(1)} \leq z_{\tau(2)}\leq\ldots\leq z_{\tau(n)}\leq\infty$ on the disk boundary.

The conjecture (\ref{3.9}) of \cite{Stieberger:2014hba} is based on equivalent conjectures on an $(n{-}3)! \times (n{-}3)!$ 
basis of disk and sphere integrals that have been made in \cite{Schlotterer:2012ny, Stieberger:2013wea}.
The latter conjectures are based on an experimental order-by-order inspection of the output of the KLT relations (\ref{2.16}), 
e.g.\ up to transcendental weight 18 at five points or weight 9 at six points:
The MZVs in the $\alpha'$-expansions of sphere integrals were observed to realize the representation (\ref{svf}) of the single-valued map in the $f$-alphabet when comparing
with the dependence of disk integrals on $s_{ij}$ \cite{Schlotterer:2012ny, Stieberger:2013wea}.
Assuming that (\ref{3.9}) holds for said $(n{-}3)! \times (n{-}3)!$ bases of disk and sphere integrals, 
integration-by-parts and monodromy relations (\ref{2.13}) and (\ref{2.14}) imply its general validity for 
arbitrary pairs of permutations $\tau,\rho \in S_{n} $ \cite{Stieberger:2014hba}. As emphasized in the reference, 
this argument relies on the action of the sv map on the trigonometric functions ${\rm sv} \,  \sin(\pi s_{ij}) /\pi  = s_{ij} $
in the monodromy relations.

Reducing the sphere integral $J(\tau | \rho) $ to a single-valued disk integral 
has both a conceptual and a practical advantage over the KLT formula:
The low-energy expansion of (\ref{3.9}) bypasses the
spurious appearance of MZVs beyond $\zeta_{\rm sv}(n_1,\ldots, n_r)$,
and the summation over $(n{-}3)! \times (n-3)!$ terms\footnote{The KLT formula (\ref{2.16}) may also be 
rewritten more compactly with $(n{-}3)! \big(\lceil \frac{n}{2} \rceil {-} 2 \big)!\big(\lfloor \frac{n}{2} \rfloor {-} 1 \big)!$ 
terms \cite{Bern:1998sv, BjerrumBohr:2010hn}.} on the right-hand side of (\ref{2.16}) is replaced 
by a single term ${\rm sv} \, Z(\tau | \rho)$. 
The implications of (\ref{3.9}) on the leading low-energy orders of
disk and sphere integrals are spelled out in (\ref{Lau01}) and~(\ref{Lau02}).


\subsection{The proof}
\label{sec:3.3}

As the main result of this work, this section is dedicated to a proof of (\ref{3.9}).
We emphasize again that the proof is subject to the restrictions detailed at the end of the introduction.

For ease of notation, we assume the first slot of the integrals $Z(\tau|\rho)$ and $J(\tau|\rho)$
to comprise the identity permutation $\tau =1,2,\ldots,n$. This assumption does not cause any loss of generality since
all the other disk and sphere integrals with the same
relative permutation $\rho \circ \tau^{-1}$ can be inferred by relabellings of the subscripts $1\leq i,j\leq n$ of $s_{ij}$.
Moreover, it will be convenient to pick an ${\rm SL}_2$ frame where $(z_1,z_{n-1},z_n) \rightarrow(0,1,\infty)$,
such that
\begin{align}
Z(1,2,\ldots,n|\rho)&= (-1)^{n-3}\int\limits_{0\leq z_{2}\leq z_3\leq \ldots\leq z_{n-2}\leq1}
 {\rm d} z_2\,  {\rm d} z_3\, \ldots  \,{\rm d} z_{n-2}\,   \prod_{1\leq i<j<n}|z_{i,j}|^{s_{ij}} \,f(\rho)
 \label{3.10} \\
J(1,2,\ldots,n|\rho)&= -\frac{1}{\pi^{n-3}} \int\limits_{\mathbb C^{n-3}} \frac{ 
 {\rm d}^2 z_2\,  {\rm d}^2 z_3\, \ldots  \,{\rm d}^2 z_{n-2} }{
 \bar z_{1,2} \bar z_{2,3} \ldots \bar z_{n-3,n-2} \bar z_{n-2,n-1}
 }\,   \prod_{1\leq i<j<n} |z_{i,j}|^{2s_{ij}} \,f(\rho) \, .
 \label{3.11}
\end{align}
The form of the meromorphic integrand
\begin{equation}
f(\rho) := \lim_{z_n \rightarrow \infty} \frac{z_n^2}{z_{\rho(1),\rho(2)} z_{\rho(2),\rho(3)}
\ldots z_{\rho(n-1),\rho(n)}z_{\rho(n),\rho(1)}}  \, ,
\label{3.12}
\end{equation}
does not affect the subsequent arguments.
The values $\bar z_1=0$ and $\bar z_{n-1}=1$ are meant to be inserted in the denominator of (\ref{3.11}) and
subsequent expressions.

The integrals $Z$ (open string) and $J$ (closed string) are connected by a Betti-deRham duality 
\cite{Beil,SchnetzTalk}: In (\ref{3.10}) the chain of integration is bounded by the
identities $z_{i}=z_{i+1}$ for $i=1,\ldots,n{-}2$. Likewise, the integrand in (\ref{3.11}) has the 
anti-meromorphic singular divisor $\cup_{i=1}^{n-2}\{\zz_{i}=\zz_{i+1}\}$ which is the deRham version of the chain 
of integration in (\ref{3.10}). Accordingly, $J(\tau|\rho)$ becomes the deRham analogue of $Z(\tau|\rho)$. It is explained 
in \cite{Bsv} that single-valued MZVs are evaluations of deRham periods (after a projection from motivic periods into deRham periods which suppresses 
$2\pi\ii$, see also \cite{BrownMotPer}). So, it is natural that $J$ is the image of $Z$ under
the single-valued map. These statements, however, do not have the status of a theorem so that we need a proof 
of the result (\ref{3.9}).
\begin{proof}[Proof of (\ref{3.9})]
We will iteratively integrate (\ref{3.10}) and (\ref{3.11}) over the variables $z_{2},z_3,\ldots,z_{n-2}$. Let 
$Z_i(z_{ i+1},$ $\ldots,$ $z_{n-2})$ and $J_i(z_{i+1},\ldots,z_{n-2})$ denote the result after the $(i{-}1)$st integration, i.e.
\begin{align}
Z_i(z_{ i+1},\ldots,z_{n-2})&:= (-1)^{n-3} \int\limits_{0\leq z_{2}\leq z_3 \ldots\leq z_{i}\leq z_{i+1}}
 {\rm d} z_{2}\,{\rm d} z_{3}\,  \ldots  \,{\rm d} z_{i}\,   \prod_{1\leq a<b<n} |z_{a,b}|^{s_{ab}} \,f(\rho)
 \label{3.14} \\
J_i(z_{ i+1},\ldots,z_{n-2})&:= -\frac{1}{\pi^{n-3}} \int\limits_{\mathbb C^{i-1}}
\frac{  {\rm d}^2 z_2\,  {\rm d}^2 z_3\, \ldots  \,{\rm d}^2 z_{i} }{
\bar z_{1,2} \bar z_{2,3} \ldots \bar z_{n-3,n-2} \bar z_{n-2,n-1} 
}  \,  \prod_{1 \leq a<b<n} |z_{a,b}|^{2s_{ab}} \,f(\rho)    \, .
 \label{3.15}
\end{align}
The functions $Z_1(z_2,\ldots,z_{n-2})$ and $J_1(z_2,\ldots,z_{n-2})$ 
at $i=1$ are given by the integrands of (\ref{3.10}) and (\ref{3.11}), respectively.

We will show by induction that
\begin{equation}
\label{svZJi}
J_i(z_{ i+1},\ldots,z_{n-2})=\frac{(-1)^{n-i+1}\text{sv}\,Z_i(z_{ i+1},\ldots,z_{n-2})}{\pi^{n-2-i}\zz_{ 1, i+1}\zz_{ i+1, i+2}\ldots \zz_{ n-2, n-1}}
\end{equation}
for all $i=1,\ldots,n{-}2$. Because $Z(1,2,\ldots,n|\rho)=Z_{n-2}(\emptyset)$ and 
$J(1,2,\ldots,n|\rho)=J_{n-2}(\emptyset)$, this implies the theorem (\ref{3.9}) (because $\zz_{1,n-1}=-1$).

Note that the absolute values in the numerators of (\ref{3.14}) and (\ref{3.15}) 
play completely different roles in both cases. In $Z_1$ there exist no complex conjugate 
variables and $|z_{a,b}|$ with $a<b$ is $-z_{a,b}$. In fact, the only motivation for employing the absolute values
for disk integrals stems from (\ref{poles01}), where the integrand does not need any explicit reference to the permutation
$\tau$ of the integration cycle. We consider the numerator as a generating series of logarithms
with the expansion parameters $s_{ab}$. In $J_1$ the numerator is a generating series of logarithms 
in $|z_{a,b}|^2=z_{a,b}\zz_{a,b}$. Because
\begin{equation}
\text{sv}\,\log(x-y)=\log[(x-y)(\overline{x}-\overline{y})]
\label{discussionbelow}
\end{equation}
for any complex numbers or variables $x,y$,
equation (\ref{svZJi}) holds for $i=1$.

Now, assume (\ref{svZJi}) holds for $i$. In the calculation of $Z_{i+1}$, the integrand may have 
a singularity at $z_{i+1}=z_{i+2}$ or at $z_{i+1}=0$. In these cases, one has to subtract the asymptotic expansion 
at the singular locus which will be exemplified in appendix \ref{sec:4}.

Note that F.~Brown and C.~Dupont give a full mathematical proof in \cite{BD} that the subtraction of singularities is always possible.

The subtraction at $z_{i+1}=z_{i+2}$ is of the form 
$c|z_{i+1,i+2}|^{s-1}$ for some $c=c(z_{i+2},\ldots,z_{n-2})$ which is constant in $z_{i+1}$ but may depend on the
integration variables $z_{i+2},\ldots,z_{n-2}$ of later steps. The exponent in $|z_{i+1,i+2}|^{s-1}$
refers to a sum $s=\sum s_{ab}$ for some pairs $a,b$ that are determined by previous integration steps. 
Assuming that\footnote{Negative values of $s$
can be addressed via analytic continuation, based on the same form of the primitive that arises for $s>0$.} $s>0$,
the subtraction can trivially be integrated
from $0=z_1$ to $z_{i+2}$ yielding $-\frac{c}{s}\cdot |z_{1,i+2}|^s$ (providing a pole in $s=0$).
The analogous result holds for a singularity at $z_{i+1}=0$.

The systematics of the kinematic poles of disk integrals generated in this way have been discussed in the literature 
from various perspectives \cite{Mafra:2011nw, Broedel:2013tta, Mafra:2016mcc}. Note that for the present proof, we only 
need the existence of such a subtraction scheme, i.e.\ the four- and five-point examples in appendix \ref{sec:4} are 
merely displayed for illustrative purposes. The closed-string analogues of the disk integrals with kinematic poles 
can be addressed with almost identical subtraction schemes, where the primitives involve factors
of $|z_{i+1,i+2}|^{2s}$ rather than $|z_{i+1,i+2}|^{s}$. All the intermediate steps of the open-string and closed-string
subtraction scheme are related through the sv map as one can
see from the Taylor expansions of $|z_{i+1,i+2}|^{2s}$ and $|z_{i+1,i+2}|^{s}$.

After the subtraction, the integrands of (\ref{3.14}) and (\ref{3.15}) have an integrable expansion at $s_{ij}=0$ 
and we can consider the integrand as a generating series in the $s_{ij}$. With this prescription we define 
the primitive $F_i$ of $Z_i$ with respect to $z_{i+1}$ and obtain
\begin{equation}
Z_{i+1}=\int_0^{z_{i+2}}\dd z_{i+1} \, Z_i 
=F_i(z_{i+2})-F_i(0)\, .
\label{rhshere}
\end{equation}
In general, the right-hand side of (\ref{rhshere}) is a series of Laurent type whose coefficients 
are iterated integrals in the letters $0,1,z_{ k}$ for $k=i{+}2,\ldots,n{-}2$.

By the inductive assumption we have 
\begin{equation}
J_{i+1}:=\int_{\CC} \dd^2 z_{ i+1} \, J_i=\int_{\CC} \dd^2z_{i+1}
\frac{(-1)^{n-i+1}\text{sv}\,Z_i}{\pi^{n-2-i}\zz_{1, i+1}\zz_{i+1, i+2}\ldots \zz_{n-2,n-1}} \, .
\end{equation}
We calculate the integral with the residue theorem of 
section 2.8 in \cite{gf}. To do so we need a single-valued primitive 
of the integrand with respect to the holomorphic variable $z_{i+1}$. By single-valued integration
 -- see property (iii) in section \ref{sec:3.1} -- this primitive is
\begin{equation}
\mathcal{F}_i:=\frac{(-1)^{n-i+1}\text{sv}\,F_i}{\pi^{n-2-i}\zz_{1, i+1}\zz_{i+1, i+2}\ldots \zz_{ n-2, n-1}}\, .
\end{equation}
Because the denominator of $\mathcal{F}_i$ is of degree two in $\zz_{i+1}$, its anti-residue at infinity 
(the residue with respect to the anti-holomorphic variable $\zz_{ i+1}$)
vanishes. Moreover, $\mathcal{F}_i$ has simple poles at $\zz_{i+1}=\zz_{1}=0$ and at $\zz_{i+1}=\zz_{i+2}$ 
whose anti-residues are obtained by substitution.
From the residue theorem in \cite{gf} (using Stokes' theorem)\footnote{Schematically, after using Stokes' theorem we use the residue theorem in the following way
in passing to the second line of (\ref{3.residue})
$$
\oint_{\partial (\CC \setminus \{z_a,z_c\})} {\rm d}\bar z_b \frac{ f(z_b) }{\bar z_{ab} \bar z_{bc}} = -\frac{2\pi i }{\bar z_{ac}} \big( f(z_c) - f(z_a) \big)\, ,
$$
where the function $f$ is regular at $z_b=z_a,z_c$.
Note that the `boundary' of $\CC \setminus \{z_a,z_c\}$ has negative orientation. A proof of this identity is in \cite{gf}, see theorem 2.29.} we obtain
\begin{align}
J_{i+1}&= \int_\CC \dd^2 z_{ i+1} \, \frac{ \partial }{\partial z_{i+1}} \, \mathcal{F}_{i}  \notag
\\
&= \frac{(-2\pi i) }{2i} \frac{(-1)^{n-i+1}[(\text{sv}\,F_i)(z_{i+2})-(\text{sv}\,F_i)(0)]}{\pi^{n-2-i}\zz_{1,i+2}\ldots \zz_{n-2, n-1}} 
\label{3.residue} \\
&=\frac{(-1)^{n-i} \, \text{sv}\, Z_{i+1} }{\pi^{n-3-i}\zz_{1,i+2}\ldots \zz_{n-2, n-1}} 
\, . \notag
\end{align}
Because the evaluation of $F_i$ commutes with the sv-map -- see property (ii) in section \ref{sec:3.1} -- this reproduces the shifted form
$i \rightarrow i{+}1$ of the inductive assumption (\ref{svZJi}) and therefore completes the induction.
\end{proof}

The proof confirms the result of \cite{Tera:2002, Brown:2009qja} that the Laurent series of $Z$ has MZV coefficients
and provides a method to calculate them which closely follows the lines of \cite{Broedel:2013tta, Mafra:2016mcc}.
At the same time, it clarifies that the coefficients of $J$ are single-valued MZVs
which can be inferred from open-string results on $Z$ without any reference to KLT 
relations (\ref{2.16}).


\section{Conclusions}
\label{sec:5}

In this work, we have proved that the moduli-space integrals in $n$-point tree-level amplitudes of
open and closed strings are related by the sv map, confirming the conjectures of
\cite{Schlotterer:2012ny, Stieberger:2013wea, Stieberger:2014hba}. More precisely, sphere 
integrals are expressed as single-valued disk integrals, where the singular parts of the anti-meromorphic sphere 
integrand are traded for an integration cycle on the disk boundary related by Betti-deRham duality.
Our proof puts an intriguing web of connections between low-energy interactions of gauge- and gravity
states in different string theories \cite{Stieberger:2014hba, Schlotterer:2016cxa, Azevedo:2018dgo} on 
firm grounds. These results go beyond the reach of the KLT relations (\ref{2.16}) as well as the known string dualities
\cite{Hull:1994ys, Polchinski:1995df, Witten:1995ex} and call for various directions of follow-up research.

In the same way as the notion of a single-valued map applies to a variety of periods \cite{BrownMotPer},
the sv relations between string tree-level amplitudes should have an echo at loop level. At genus one,
this gives rise to expect a relation between elliptic multiple zeta values \cite{Enriquez:eMZV} in open-string
$\alpha'$-expansions \cite{Broedel:2014vla, Broedel:2017jdo} and modular graph 
functions in closed-string expansions \cite{Green:1999pv, Green:2008uj, DHoker:2015gmr, DHoker:2015wxz}\footnote{Modular 
graph functions are believed to fall into the more general framework of non-holomorphic modular forms described in \cite{Brown:nonholA, Brown:nonholB}.}. 

Single-valued polylogarithms and MZVs were found to play a key role in one-loop amplitudes of closed 
superstrings \cite{Zerbini:2015rss, DHoker:2015wxz}. Moreover, first explicit connections between open- and 
closed-string results at genus one were established in \cite{Broedel:2018izr}, along with an empirically motivated
conjecture for the form of an elliptic single-valued map. Since the proof of this
work only relies on general properties of the genus-zero integrals -- such as singularities of the
integrands and the existence of suitable primitives -- it is conceivable that similar methods can be applied to
timely research problems at genus one and beyond.

At higher genus, the $\alpha'$-expansion of moduli-space integrals of closed strings was 
pioneered in \cite{DHoker:2013fcx, DHoker:2014oxd}, and the last months witnessed tremendous progress
in understanding their systematics and degenerations \cite{DHoker:2017pvk, DHoker:2018mys}. However, a 
higher-genus framework of elliptic multiple zeta values is still lacking, so the knowledge of open-string low-energy 
expansions is very limited. We hope that the ideas of the proof in this work 
are helpful to identify a language for loop-level integrals in open- and closed-string amplitudes
that is tailored to expose their relations.


\section*{Acknowledgements}

We are grateful to the Hausdorff Institute Bonn for providing stimulating atmosphere, support, and hospitality through the Hausdorff Trimester Program ``Periods in Number Theory, Algebraic Geometry
and Physics'' and the workshop ``Amplitudes and Periods'' where this work was initiated. 
Moreover, we would like to thank Francis Brown, Pierre Vanhove and Federico Zerbini for inspiring discussions.
The research of Oliver Schlotterer was supported in part by Perimeter Institute for Theoretical Physics. Research at Perimeter Institute is supported by the Government of Canada through
the Department of Innovation, Science and Economic Development Canada and by the Province of Ontario through the Ministry of Research, Innovation and Science.
Oliver Schnetz is supported by DFG grant SCHN~1240/2.


\appendix


\section{Pole subtractions}
\label{sec:4}

In this appendix, we illustrate the subtraction of singularities in the successive integration over
disk punctures, see the discussion below (\ref{discussionbelow}). In the representation 
(\ref{3.10}) of disk integrals, the rational function $f(\rho)$ defined in (\ref{3.12})
may contribute a pole in $z_{i+1,i+2}$ or $z_{i+1}$ to the
integrand of $\int^{z_{i+2}}_0 \dd z_{i+1}$ in the induction step of the main proof.
An explicit realization of subtraction schemes will now be
spelled out for certain four- and five-point integrals which reflect the key features of the strategy at $n$ points.
Still, we reiterate that the proof in section \ref{sec:3.3} only requires the existence of a subtraction scheme,
i.e.\ the details of the subsequent examples are just given to illustrate the general mechanism. 

Similar subtractions were done in the more complicated framework of $\phi^4$ quantum field theory 
in \cite{numfunct} to obtain the seven loop beta-function (see Figure 7 and Conjecture 4.12). In tree-level
amplitudes of string theories, the singularities are logarithmic once the disk and sphere integrals are brought
into the form of $Z(\tau|\rho)$ and  $J(\tau|\rho)$ via integration by parts. Since there is no need for dimensional 
regularization in string tree-level amplitudes, the analogue of Conjecture 4.12 in \cite{numfunct}
becomes a lemma that follows from blowing up all singular loci in the integrand.
See \cite{BEK} for the application of the concept of blowing up singularities in the context of quantum field theory.


\subsection{Four-point examples}
\label{sec:4.1}

In an ${\rm SL}_2(\mathbb R)$ frame with $(z_1,z_3,z_4) \rightarrow (0,1,\infty)$, we consider the following instances of
the disk and sphere integrals (\ref{3.10}) and (\ref{3.11}) with a single kinematic pole,
\begin{align}
Z(1,2,3,4|1,2,4,3) &= - \int^1_0 \dd z_2 \, \frac{ z_2^{s_{12}}  (1{-}z_2)^{s_{23}} }{z_2} =  - \int^1_0 \dd z_2 \, \frac{ z_2^{s_{12}} }{z_2}  \, \big( \underbrace{ (1{-}z_2)^{s_{23}} - 1}_{(i)} +\underbrace{ 1 }_{(ii)}\big) \label{poles03} 
\\
Z(1,2,3,4|1,4,2,3) &= - \int^1_0 \dd z_2 \, \frac{ z_2^{s_{12}}  (1{-}z_2)^{s_{23}} }{1-z_2} =  - \int^1_0 \dd z_2 \, \frac{ (1{-}z_2)^{s_{23}}   }{1-z_2}  \, \big( \underbrace{z_2^{s_{12}} - 1}_{(iii)} +\underbrace{ 1}_{(iv)} \big) \label{poles04} 
\\
J(1,2,3,4|1,2,4,3) &=  \int\limits_{\mathbb C}  \frac{ \dd^2 z_2 \, |z_2 |^{2s_{12}}  |1{-}z_2 |^{2s_{23}} }{\pi \, z_2 \, \bar z_2 (\bar z_2-1)} 
=  \int\limits_{\mathbb C}  \frac{ \dd^2 z_2 \, |z_2 |^{2s_{12}} \big( \overbrace{ |1{-}z_2 |^{2s_{23}} - 1}^{(v)} +\overbrace{ 1 }^{(vi)}\big) }{\pi \, z_2 \, \bar z_2 (\bar z_2-1)} 
\label{poles05} 
\\
J(1,2,3,4|1,4,2,3) &=  
\int\limits_{\mathbb C}  \frac{ \dd^2 z_2 \, |z_2 |^{2s_{12}}  |1{-}z_2 |^{2s_{23}} }{\pi \, (1-z_2) \, \bar z_2 (\bar z_2-1)} 
=  \int\limits_{\mathbb C}  \frac{ \dd^2 z_2 \,  |1{-}z_2 |^{2s_{23}}    \big( \overbrace{  |z_2 |^{2s_{12}} - 1}^{(vii)} +\overbrace{ 1 }^{(viii)}\big) }{\pi \, (1-z_2) \, \bar z_2 (\bar z_2-1)}  \, ,
 \label{poles06} 
\end{align}
where the shorthands $(i)$ to $(viii)$ refer to the full-fledged integrals after isolating the
highlighted terms in the sums $(\ldots)$ of the integrand, e.g.
\begin{equation}
(v)=  \int\limits_{\mathbb C}  \frac{ \dd^2 z_2 \, |z_2 |^{2s_{12}} \big(  |1{-}z_2 |^{2s_{23}} - 1 \big) }{\pi \, z_2 \, \bar z_2 (\bar z_2-1)} 
\, .
\end{equation}
The subtractions on the right-hand side are tailored to isolate the field-theory limits
\begin{align}
Z(1,2,3,4|1,2,4,3)  &=- \frac{1}{s_{12}} + O(\alpha') \ , \ \ \ \
J(1,2,3,4|1,2,4,3)  =- \frac{1}{s_{12}} + O(\alpha') \label{poles07} \\
Z(1,2,3,4|1,4,2,3) & =- \frac{1}{s_{23}} + O(\alpha') \ , \ \ \ \
J(1,2,3,4|1,4,2,3)  =- \frac{1}{s_{23}} + O(\alpha') \ ,
\label{poles08}
\end{align}
which can be straightforwardly generated from the integrals
\begin{align}
(ii) &= -\int^1_0 \dd z_2 \, z_2^{s_{12}-1}
= - \frac{z_2^{s_{12}}}{s_{12}} \, \Big|^{z_{2}=1}_{z_2=0} = - \frac{1}{s_{12}} 
\label{poles09} \\
(iv) &= -\int^1_0 \dd z_2 \, (1-z_2)^{s_{23}-1} = \frac{(1-z_2)^{s_{23}}}{s_{23}} \, \Big|^{z_{2}=1}_{z_2=0} = - \frac{1}{s_{23}} 
\label{poles10} \\
(vi) &= \int\limits_{\CC\setminus \{0,1\}} \frac{ \dd^2 z_2 \, |z_2|^{2s_{12}} }{\pi \, z_2 \, \bar z_2 \,(\bar z_2-1) } =
\frac{1}{\pi s_{12}}  \int\limits_{\CC\setminus \{0,1\}} \dd^2 z_2 \, \frac{\partial}{\partial z_2} \, \frac{  |z_2|^{2s_{12}} }{\bar z_2 \,(\bar z_2-1) } 
\notag \\
&= \frac{1}{2\pi i\, s_{12}}  \oint\limits_{\partial (\mathbb C \setminus \{0,1\})}   \frac{\dd \bar z_2 \,  |z_2|^{2s_{12}} }{\bar z_2 \,(\bar z_2-1) } \label{poles11} \\
&=\frac{1}{2\pi i\, s_{12}} \Big\{ -\frac{2\pi i \, |z_2|^{2s_{12} } }{\bar z_2 - 1} \, \Big|_{z_2=0}
-  \frac{2\pi i \, |z_2|^{2s_{12} } }{\bar z_2 } \, \Big|_{z_2=1} \Big\} = -\frac{1}{s_{12}} \ . \notag
\end{align}
The evaluation of $(viii)$ is completely analogous to $(vi)$ and yields $-1/s_{23}$. Note that, following the proof in
section \ref{sec:3.3}, the meromorphic parts of the primitives in (\ref{poles09}) and (\ref{poles11}) are identical.
So, the fact that $(vi) = {\rm sv} \,(ii)$ is clear from the general arguments given above and confirmed by the inspection
of the final result $-\frac{1}{s_{12}}$ in both cases, where the action of sv trivializes.

The integrands in the curly bracket of $(i), (v)$ and $(iii), (vii)$ are designed to
be regular as $z_2 \rightarrow 0$ and $z_2\rightarrow 1$, respectively. This renders the 
integrated expressions non-singular w.r.t.\ $s_{ij}$, and
the arguments in the proof in section \ref{sec:3.3} can be applied to the series in $\log (z_{ij})$ and $\log|z_{ij}|^2$ 
without the need for further subtractions: Along with each monomial in $s_{12}^m s_{23}^n$ with $m,n\geq 0$, 
the holomorphic primitives of $(\log z_2)^m  (\log (1{-}z_2))^n/z_2$ and $(\log |z_2|^2)^m  (\log |1{-}z_2|^2)^n/z_2$ 
are related by the sv map and ultimately evaluated at $z_2=1$. Hence, at the level of the resulting MZVs, 
\begin{equation}
(v) = {\rm sv} \,(i) \ , \ \ \ \ \ \ 
(vii) = {\rm sv} \,(iii) \ .
\label{sv4pt}
\end{equation}


\subsection{Five-point examples: non-overlapping singularities}
\label{sec:4.2}

Starting from five-point disk and sphere integrals, the residues of the kinematic poles are
by themselves series in $s_{ij}$ with MZV coefficients. As a first example, we consider the integral
\begin{align}
Z(1,2,3,4,5 | 1,2,5,3,4) &= \int^1_0 \dd z_3 \int^{z_3}_0 \dd z_2 \, \frac{ z_2^{s_{12}} z_3^{s_{13}} z_{32}^{s_{23}} (1 {-}z_2 )^{s_{24} } (1{-}z_3)^{s_{34}} }{z_2 \, (z_3{-}1)} \notag \\
&= - \frac{1}{s_{12} s_{34}} + O(\alpha'^0) \label{poles12}
\end{align}
in an ${\rm SL}_2(\mathbb R)$ frame with $(z_1,z_4,z_5) \rightarrow (0,1,\infty)$, where the
poles $s_{12}^{-1}$ and $s_{34}^{-1}$ stem from different endpoints $z_2 \rightarrow 0$ and $z_3\rightarrow 1$
of the integration domain $0\leq z_2 \leq z_3\leq 1$. As an analogue of the
subtraction scheme in (\ref{poles03}) to (\ref{poles06}), we rewrite the integrand of (\ref{poles12}) as
\begin{align}
&Z(1,2,3,4,5 | 1,2,5,3,4) = \int^1_0 \dd z_3 \, \frac{   z_3^{s_{13}}  (1{-}z_3)^{s_{34}} }{z_3{-}1} \notag \\
& \ \ \ \ \ \int^{z_3}_0 \dd z_2 \, \frac{ z_2^{s_{12}}   }{z_2 } \, \big( \underbrace{  z_{32}^{s_{23}} (1{-}z_2 )^{s_{24} } - z_3^{s_{23}} }_{(\alpha)}+\underbrace{ z_3^{s_{23}} }_{(\beta)} \big) \ .
\label{poles13}
\end{align}
The contribution of $(\beta)$ involves a straightforward integral over $z_2$ similar to (\ref{poles09}) along with 
an integral over $z_3$ of four-point type, cf.\ (\ref{poles04}),
\begin{align}
(\beta) &=\int^1_0 \dd z_3 \, \frac{   z_3^{s_{13}+s_{23}}  (1{-}z_3)^{s_{34}} }{(z_3{-}1)} \frac{ z_2^{s_{12}} }{s_{12}} \Big|^{z_2 = z_3}_{z_2=0}  \notag \\
&= \frac{1}{s_{12}} \int^1_0 \dd z_3 \, \frac{   z_3^{s_{12}+s_{13}+s_{23}}  (1{-}z_3)^{s_{34}} }{z_3{-}1}
\label{poles14} \\
&=  \frac{1}{s_{12}} \Big( Z(1,2,3,4|1,4,2,3) \Big|^{s_{23} \rightarrow s_{34}}_{s_{12} \rightarrow s_{12}+s_{13}+s_{23}} \Big) \notag \ .
\end{align}
The leading order of $Z(1,2,3,4|1,4,2,3)$ in (\ref{poles08}) then yields the low-energy limit $-(s_{12} s_{34})^{-1}$ of
the integral $Z(1,2,3,4,5 | 1,2,5,3,4)$, see (\ref{poles12}).

The integrand of the contribution of $(\alpha)$ in (\ref{poles13})
is regular at $z_2=0$, so the integral over $z_2$ 
\begin{align}
H(z_3) &:= \int^{z_3}_0 \dd z_2 \, \frac{ z_2^{s_{12}} }{z_{2}} \big(  z_{32}^{s_{23}} (1{-}z_2)^{s_{24} } - z_3^{s_{23}}  \big)
\label{poles15}
\end{align}
does not involve any singularity in $s_{ij}$, and the $\alpha'$-expansion can be performed at the
level of the $\log(z_{ij})$ in the integrand. The limit  $z_3\rightarrow 1$ of (\ref{poles15}) is smooth
and again reproduces an integral of four-point type, cf.\ $(i)$ in (\ref{poles03})
\begin{align}
H(1) &= \int^{1}_0 \dd z_2 \, \frac{ z_2^{s_{12}} }{z_{2}} \big(  (1{-}z_2 )^{s_{23}+s_{24} } - 1  \big) \, .
\label{poles16}
\end{align}
The contribution of $(\alpha)$ in (\ref{poles13}) still yields a pole in $s_{34}$
upon integration over $z_3$. This pole can be traced back to the factor of 
$\frac{    (1{-}z_3)^{s_{34}} }{(z_3{-}1)}$, and we isolate it by the subtraction scheme
\begin{align}
(\alpha) = \int^1_0 \dd z_3 \, \frac{    (1{-}z_3)^{s_{34}} }{z_3{-}1} \, \big( \underbrace{ z_3^{s_{13}} H(z_3)  - H(1) }_{(\gamma)}+\underbrace{ H(1) }_{(\delta)} \big) \, .
\label{poles17}
\end{align}
The integral in (\ref{poles10}) determines
\begin{equation}
(\delta) = -\frac{H(1) }{s_{34}} \ ,
\label{poles18}
\end{equation}
and the integrand for the contribution $(\gamma)$ to (\ref{poles17})
is regular at $z_3 \rightarrow 1$ such that
\begin{equation}
(\gamma)= \int^1_0 \dd z_3 \, \frac{    (1{-}z_3)^{s_{34}} }{z_3{-}1} \, \big( z_3^{s_{13}} H(z_3)  - H(1)  \big) 
\label{poles19}
\end{equation}
is regular in $s_{34}$ and can be $\alpha'$-expanded at the level of the integrand.

In adapting the subtraction scheme to the corresponding sphere integral
\begin{align}
&J(1,2,3,4,5 | 1,2,5,3,4) = -\frac{1}{\pi^2}\int_{\CC^2}  \frac{\dd^2 z_2 \,\dd^2 z_3}{\bar z_{12} \bar z_{23} \bar z_{34}}  \frac{   |z_3|^{2s_{13}}  |1{-}z_3|^{2s_{34}} }{(z_3{-}1)} \notag \\
& \ \ \ \   \ \ \ \   \ \ \ \  \times \frac{ |z_2|^{2s_{12}}   }{z_2 } \, \big( \underbrace{  |z_{23}|^{2s_{23}} |1{-}z_2 |^{2s_{24} } - |z_3|^{2s_{23}} }_{(A)}+\underbrace{ |z_3|^{2s_{23}} }_{(B)} \big) \ ,
\label{poles93}
\end{align}
the primitives for all contributions $(\alpha), (\beta), (\gamma), (\delta)$ have the same 
meromorphic parts as in the case of $J(1,2,3,4,5 | 1,2,5,3,4) $. In analogy 
with (\ref{poles14}), we have
\begin{align}
(B) &= -\frac{1}{2\ii \pi^2 s_{12}} \int_{\CC} \dd^2 z_3 \oint_{
\text{`}\partial (\CC\setminus\{0,z_3\})\text{'}} \dd \bar z_2 \ \frac{   |z_2|^{2s_{12}}  |z_3|^{2s_{13}+2s_{23}}  |1{-}z_3|^{2s_{34}} }{\bar z_{12} \bar z_{23} \bar z_{34} \, (z_3{-}1)} \notag \\
&= \frac{1}{\pi s_{12}} \int_{\CC} \dd^2 z_3 \, \frac{   |z_3|^{2(s_{12}+s_{13}+s_{23})}  |1{-}z_3|^{2s_{34}} }{\bar z_{13} \bar z_{34}\,(z_3{-}1)}
\label{poles94} \\
&=  \frac{1}{s_{12}} \Big( J(1,2,3,4|1,4,2,3) \Big|^{s_{23} \rightarrow s_{34}}_{s_{12} \rightarrow s_{12}+s_{13}+s_{23}} \Big) \notag \ ,
\end{align}
which gives the desired expression ${\rm sv}\, (\beta)$.

The $z_2$ integral of $(A)$,
\begin{equation}
I(z_3):=\frac{1}{\pi}\int_{\CC}\dd^2 z_2 \, \frac{ |z_2|^{2s_{12}}   }{z_2 \, \bar{z}_{12} \bar{z}_{23}}\big(|z_{23}|^{2s_{23}} |1{-}z_2 |^{2s_{24}}-|z_3|^{2s_{23}}\big)
\label{this1}
\end{equation}
is regular and the general method in the proof of the main result applies. We obtain:
\begin{equation}
I(z_3) =-\frac{1}{\zz_{13}}\text{sv}\,H(z_3) \, .
\label{this2}
\end{equation}
Upon insertion into (\ref{poles93}), this implies
\begin{equation}
(A)=\frac{1}{\pi}\int_{\CC}\frac{\dd^2 z_3}{\bar z_{13} \bar z_{34}}  \frac{ |1{-}z_3|^{2s_{34}} }{(z_3{-}1)} \big(\underbrace{ |z_3|^{2s_{13}}\text{sv}\,H(z_3)-\text{sv}\,H(1)}_{(C)}
+\underbrace{\text{sv}\,H(1)}_{(D)}\big) \, .
\label{hereC}
\end{equation}
In analogy to (\ref{poles11}) the integral in the last term gives 
\begin{equation}
(D)=-\frac{\text{sv}\,H(1)}{s_{34}} \, ,
\label{changesign}
\end{equation}
which is identical to $ {\rm sv} \, (\delta)$ by (\ref{poles18}).
The integral $(C)$ in (\ref{hereC}) is regular and can be expanded in $\alpha'$ in the integrand.
By the general method in the proof of the main result, we obtain $(C)={\rm sv} \, (\gamma)$
and recover $J(1,2,3,4,5 | 1,2,5,3,4) = {\rm sv} \, Z(1,2,3,4,5 | 1,2,5,3,4) $ term by term
in the subtraction scheme.


\subsection{Five-point examples: nested singularities}
\label{sec:4.3}

While the singularities of the five-point example in appendix \ref{sec:4.2} stem from different regions
$z_2\rightarrow 0$ and $z_3\rightarrow 1$, the following disk integral acquires 
kinematic poles in $s_{123} := s_{12}+s_{13}+s_{23}$ from the nested singularity\footnote{In a five-point setup, one can still avoid the nested singularities by representing (\ref{poles112})
in a different ${\rm SL}_2$ frame, but this is no longer true at six points. We choose the ${\rm SL}_2$ frame
with $(z_1,z_4,z_5) \rightarrow (0,1,\infty)$ here to illustrate that the nesting of singularities does not 
obstruct the existence of a subtraction scheme.} in the integration region where $z_2,z_3\rightarrow 0$:
\begin{align}
Z_{\rm nest} &= - Z(1,2,3,4,5 | 1,2,3,5,4) -Z(1,2,3,4,5 | 1,3,2,5,4) 
\notag \\
&= \int^1_0 \dd z_3 \int^{z_3}_0 \dd z_2 \, \frac{ z_2^{s_{12}} z_3^{s_{13}} z_{32}^{s_{23}} (1{-}z_2 )^{s_{24} } (1{-}z_3)^{s_{34}} }{z_2 \, z_3} \label{poles112} \\
&= \frac{1}{s_{12} s_{123}} + O(\alpha'^0) \, . \notag
\end{align}
The first step of the subtraction scheme closely follows the lines of (\ref{poles13})
\begin{align}
Z_{\rm nest} &=  \int^1_0 \dd z_3 \frac{ z_3^{s_{13}}  (1{-}z_3)^{s_{34}}  }{z_3}
 \int^{z_3}_0 \dd z_2 \, \frac{ z_2^{s_{12}} }{z_2} \big(  \underbrace{  z_{32}^{s_{23}}(1{-}z_2)^{s_{24}}  -  z_3^{s_{23}} }_{(p) } + \underbrace{ z_3^{s_{23}} }_{(q)} \big) \label{poles113} \, ,
\end{align}
and the evaluation of the second contribution $(q)$ is almost identical to $(\beta)$ in (\ref{poles14}),
\begin{align}
(q) &=\int^1_0 \dd z_3 \, \frac{   z_3^{s_{13}+s_{23}}  (1{-}z_3)^{s_{34}} }{z_3} \frac{ z_2^{s_{12}} }{s_{12}} \Big|^{z_2 = z_3}_{z_2=0}  =  - \frac{1}{s_{12}} \Big( Z(1,2,3,4|1,2,4,3) \Big|^{s_{23} \rightarrow s_{34}}_{s_{12} \rightarrow s_{123}} \Big) \, .
\label{poles114} 
\end{align}
The low-energy limit $Z_{\rm nest} =\frac{1}{s_{12} s_{123}} + O(\alpha'^0)$ in (\ref{poles112}) then stems from the leading term
of the four-point integral $Z(1,2,3,4|1,2,4,3)$ in (\ref{poles07}) at shifted first argument $s_{12} \rightarrow s_{123}$.

In the subtraction scheme for 
\begin{align}
(p)= \int^1_0 \dd z_3 \, \frac{ z_3^{s_{13}} (1{-}z_3)^{s_{34}} H(z_3) }{z_3}  \, , 
\label{poles117}
\end{align}
it would be tempting to closely follow the treatment of $(\alpha)$ in (\ref{poles17})
and to subtract the $z_3\rightarrow0$ limit of the quantity $H(z_3)$ in (\ref{poles15}).
However, this limit does not admit a regular $\alpha'$-expansion and we shall instead write
$H(z_3) = z_3^{s_{12}+s_{23}}h(z_3)$
(which extracts the exact scaling behavior of $H$ at $z_3=0$). 
We set $z_2=xz_3$ in the integral representation (\ref{poles15}) of $H(z_3)$ and obtain\footnote{Note that
the leading terms of the $\alpha'$-expansion of (\ref{poles15new}) are given by
\begin{align*}
h(z_3)&= s_{24} I(0,10,z_3) - s_{23} \zeta(2)  + s_{24}^2 I(0,110,z_3) + s_{23}^2 \zeta(3)  - s_{12}s_{24} I(0,100,z_3) 
\\
& \ \ \ \ \ \ \ + s_{12}s_{23} \zeta(3) + s_{23}s_{24} \big[ I(0,110,z_3) - I(0, 100,z_3) \big] + O(\alpha'^3)
\, ,
\end{align*}
see (\ref{3.1}) for the definition of the iterated integrals $I(0,a_1a_2\ldots a_w,z)$.}
\begin{align}
h(z_3) &= \int^{1}_0 \frac{ {\rm d} x}{x} \, x^{s_{12}} \big(  (1{-}x)^{s_{23}} (1{-}z_3 x )^{s_{24} } - 1 \big) \, , \ \ \ \ z_3 \leq 1\, .
\label{poles15new}
\end{align}
Since (\ref{poles15new}) is regular as
$z_3\rightarrow 0$, the appropriate analogue of (\ref{poles17}) is
\begin{equation}
(p)= \int^1_0 \dd z_3 \, \frac{  z_3^{s_{123}}  }{z_3}  \big(
\underbrace{ (1{-}z_3)^{s_{34}} h(z_3)- h(0) }_{(r)} + \underbrace{ h(0) }_{(t)}
\big) \, ,
\label{poles118}
\end{equation}
where the integrand in $(r)$ is regular as $z_3\rightarrow 0$.
The integral can be performed order by order. Finally, the pole from the nested singularity
\begin{equation}
(t)= \frac{ h(0) }{s_{123}} = \frac{1}{s_{123}} \int^1_0 \frac{\dd x}{x} \,x^{s_{12}} \big( (1{-}x)^{s_{23}} - 1 \big) \, ,
\label{poles119}
\end{equation}
has a residue identical to $(i)$ in (\ref{poles03}).

For the corresponding sphere integral
\begin{align}
J_{\rm nest} &= - J(1,2,3,4,5 | 1,2,3,5,4) -J(1,2,3,4,5 | 1,3,2,5,4) 
\notag \\
&= - \frac{1}{\pi^2} \int_{\CC^2} \frac{ \dd^2 z_2 \, \dd^2 z_3}{ \bar z_{12} \bar z_{23} \bar z_{34}}
\, \frac{ |z_2|^{2s_{12}} |z_3|^{2s_{13}} |1{-}z_3|^{2s_{34}} }{ z_2 z_3} 
\big( \underbrace{  |z_{23}|^{2s_{23}} |1{-}z_2 |^{2s_{24} } - |z_3|^{2s_{23}} }_{(P)}+\underbrace{ |z_3|^{2s_{23}} }_{(Q)} \big)\, ,
\label{poles212} 
\end{align}
the first step of the subtraction scheme is again almost identical to (\ref{poles93}), resulting in
\begin{equation}
(Q)= - \frac{1}{s_{12}} \Big( J(1,2,3,4|1,2,4,3) \Big|^{s_{23} \rightarrow s_{34}}_{s_{12} \rightarrow s_{123}} \Big) \, ,
\label{poles219}
\end{equation}
which matches ${\rm sv} \, (q)$ by (\ref{poles114}).

The $z_2$ integral of $(P)$ is again given by (\ref{this1}), and we 
will use its representation in (\ref{this2}),
\begin{equation}
(P)= \frac{1}{\pi} \int_{\mathbb C} \frac{ \dd^2 z_3}{ \bar z_{13} \bar z_{34}} \frac{ |z_3|^{2s_{13}} |1-z_3|^{2 s_{34}}  }{z_3} \, {\rm sv} \, H(z_3)\, .
\label{this3}
\end{equation}
Then, we use the single-valued analogue ${\rm sv} \, H(z_3) 
= |z_3|^{2s_{12}+2s_{23}} {\rm sv} \, h(z_3)$ of the above
rewriting $H(z_3) = z_3^{s_{12}+s_{23}}h(z_3)$ with 
$h(z_3)$ given by (\ref{poles15new}) and employ the following subtraction scheme:
\begin{equation}
(P)= \frac{1}{\pi} \int_{\mathbb C} \frac{ \dd^2 z_3}{ \bar z_{13} \bar z_{34}} \frac{ |z_3|^{2s_{123}} }{z_3} \,\big( \underbrace{ |1-z_3|^{2 s_{34}}   {\rm sv} \, h(z_3) - {\rm sv} \, h(0) }_{(R)} + \underbrace{ {\rm sv} \, h(0) }_{(T)} \big)\, .
\label{this4}
\end{equation}
The integrand in $(R)$ is regular as $z_3\rightarrow 0$ and we arrive at
$(R)={\rm sv}\,(r)$ upon order-by-order integration, cf.\ (\ref{poles118}). The last term in (\ref{this4}) 
can be trivially integrated to give
\begin{equation}
(T)= \frac{ {\rm sv} \, h(0) }{s_{123}} \, ,
\label{this5}
\end{equation}
which agrees with ${\rm sv} \, (t)$ by (\ref{poles119}). Hence, we have checked the 
relation $J_{\rm nest}= {\rm sv} \, Z_{\rm nest}$ at the level of all the terms in the
subtraction scheme.

\bibliographystyle{plain}
\renewcommand\refname{References}

\end{document}